# Robust ferroelectric properties of organic Croconic Acid films grown on spintronically relevant substrates

Sambit Mohapatra [a], Victor Da Costa [a], Garen Avedissian [a], Jacek Arabski [a], Wolfgang Weber [a], Martin Bowen [a], Samy Boukari [a]

[a] Université de Strasbourg, CNRS, Institut de Physique et Chimie des Matériaux de Strasbourg, UMR 7504, F-67000 Strasbourg, France

*Correspondence E-mail: sambit.mohapatra@ipcms.unistra.fr

**Abstract:** The discovery of stable room temperature ferroelectricity in Croconic Acid, an organic ferroelectric material, with polarization values on par with those found in inorganic ferroelectric materials and highest among organic ferroelectric materials, has opened up possibilities to realize myriads of nano-electronic and spintronic devices based on organic ferroelectrics. Such possibilities require an adequate understanding of the ferroelectric properties of Croconic Acid grown on surfaces that are commonly employed in device fabrication. While several macroscopic studies on relatively larger crystals of Croconic Acid have been performed, studies on thin films are only in their early stages. We have grown thin films of Croconic Acid on gold and cobalt surfaces, which are commonly used in spintronic devices as metallic electrodes, and studied the films' ferroelectric response using ex-situ Piezoresponse Force Microscopy at room temperature. We show that the polarization reversal in Croconic Acid domains is sensitive to the substrate surface. Using the same experimental protocol, we observe the robust polarization reversal of a single, mostly in-plane electrical domain for a cobalt substrate, whereas polarization reversal is hardly observed for a gold

**substrate. We attribute this difference to the substrate's influence on the Croconic Acid molecular networks. Our study suggests that to realize devices one has to take care about the substrate on which Croconic Acid will be deposited. The fact that polarization switching is robust on cobalt surface can be used to fabricate multifunctional devices that utilize the cobalt/Croconic Acid interface.**

Ferroelectric materials have found substantial applications in a number of fields ranging from electronic-spintronic devices and sensors-transducers to energy harvesting and heat transfer applications.[1–3] In the field of electronic devices, the use of ferroelectric materials has contributed to the realization of a plethora of devices, starting from memory devices to voltage-operated spintronic devices. Organic ferroelectric materials may provide added advantages due to their flexibility, chemical functionality, non-toxicity, low cost, easy processing and low power consumption. Organic ferroelectric materials based devices such as ferroelectric random access memory (FeRAM)[4], ferroelectric field effect transistors (FeFET)[5], ferroelectric tunnel junctions (FTJ)[6,7], ferroelectric optoelectronic devices[8], ferroelectric resistive switches[9], ferroelectric capacitors[10,11] and memory diodes[9,12,13] have been realized in the recent past.

However, the number of organic ferroelectric materials is small and potential applications of organic ferroelectric materials have been limited to polymer-based materials. Not only do polymeric materials exhibit a small polarization value compared to inorganic materials, but they are also not suitable for low-voltage applications due to the high electric field required to switch the polarization[14,15] and the presence of leakage current[16,17] in vertical devices.

Recently, organic ferroelectrics with large polarization values have been explored. For example, crystals of Croconic Acid (CA) have been found to show a room temperature polarization value as large as 30 µC/cm$^2$.[18] CA falls in the special category of hydrogen-bonded organic ferroelectrics with its ferroelectricity originating from a synchronized proton transfer mechanism with the help of the π-electrons. It results in the highest crystal polarization among organic ferroelectric materials; the polarization is even higher than several common inorganic ferroelectric materials such as $BaTiO_3$ [19, 20] or $SrBi_2Ta_2O_9$.[21] The polarization in CA can be switched with low fields at frequencies as high as 1 kHz[22], making it a potential candidate for low-voltage device operations. Furthermore, the observation of a large second-order non-linear susceptibility[23] and the predicted possible photoinduced ferroelectric switching[24] suggest that CA can be potentially useful for non-linear optical or optoelectronic devices.

Since the discovery of stable room-temperature ferroelectricity in CA[22], most work have focused on the crystals. But studies on thin films, which are necessary to develop devices, are only in their early stages. Jiang et al.[25] successfully manipulated the polarization state of CA thins films deposited on $Al_2O_3$, but did not discuss the orientation of polarization. Attempts to use CA as a ferroelectric gate in an Organic FET (OFET) were also carried out but the devices suffer from a high leakage current.[26] Finally, scanning tunneling microscopy studies of CA ultra-thin films (in a few monolayers range) have been reported.[26, 27] However, detailed microscopic studies on bulk ferroelectric properties of CA thin films on metal surfaces are yet to be performed.

The design of any spintronic or electronic device involves a crucial step of fabricating the electrode with the right material and in the right way. Usually, a metallic thin film is used as an electrode. In most cases, the interface between the metal electrode and the device plays a vital role in generating the device's electronic response. Especially, with reduction in size, electronic devices become more and more sensitive to the physical properties that emerge at the interface, where different materials interact. In the worst case, especially with molecules, their interaction with another material, e.g. with a metal, can impede the properties for which they were chosen. When organic molecules are interfaced with metallic substrates, intriguing phenomena can occur, depending on the surface reactivity.

For example, it was impossible to switch the spin state of spin crossover molecules directly deposited on metallic Cu(100)[29]. The switching properties were nevertheless restored upon passivation of the reactive surface by nitrogenation of the metallic copper. For spintronic applications, where one usually has to use more reactive metallic substrate like Co, molecules' properties could be preserved by the intercalation of a thin comparatively less reactive metal spacer like Cu, without losing the high interface polarization necessary for effects such as magnetoresistance[30].

Similarly, stabilization of hydrogen bonded crystalline two dimensional molecular network structure of Rhodizonic Acid, another hydrogen bonded molecular member of the CA family, which is believed to be necessary for possible ferroelectric ordering, is possible on Au even at elevated temperatures. On the contrary, on a comparatively more reactive Cu surface, it is observed that deprotonation of the molecule and reaction with Cu atoms can happen upon substrate annealing that leads to a different structural arrangement which may in turn influence the expected ferroelectric properties of the molecular layer. [31]

Although the above examples of substrate induced effect on the properties of organic molecular materials are confined to the case of a few monolayers, these substrate dependent variations of interface properties may also have substantial impact on the bulk properties of the material in thicker films that are present in devices. Thus, it is of utmost importance to study the substrate induced effect on the properties of organic films not only in the thin monolayer limit but also in bulk thick films. Therefore, prior to the realization of electronic or spintronic applications of CA thin films, it is essential to study the ferroelectric properties of CA on relevant metallic substrate surfaces under ambient conditions, especially on substrates with high surface reactivity. With this in mind, we have investigated the stability of the properties of bulk CA films grown on two commonly used spintronic electrodes with similar crystalline nature (supplementary figure S6) but with different surface reactivity at room temperature and under atmospheric pressure.

We studied the microscopic ferroelectric properties of CA thin films deposited on gold (Au) and comparatively more reactive cobalt (Co) surfaces which are relevant to fabricate spintronic devices[32]. Piezoresponse Force Microscopy (PFM) results demonstrate that the ability to switch the polarization state of a domain within the CA film depends on which underlying substrate is used. We find that robust and reproducible switching can occur for a Co substrate, but not for an Au substrate which is normally less reactive than the Co surface. The fact that robust ferroelectric properties of CA can be stabilized on Co surface without the need for any intercalating layer is quite interesting for spintronic applications employing such a hybrid interface.

A thin films (~50-150 nm) were deposited in an ultra-high vacuum (UHV) chamber on Au and Co substrates with a base pressure of ~$5\times10^{-10}$ mbar by the thermal evaporation of commercially available CA powder from an effusion cell operating at a temperature of ~130 °C. The pressure during deposition was ~$1\times10^{-8}$ mbar. The substrates were kept at room temperature and rotated uniformly to reduce thickness gradients. No capping layer was deposited.

To obtain an Au substrate, we deposited 150 nm of Au by Molecular beam epitaxy in a separate UHV chamber onto freshly cleaved mica. These mica/Au substrates provide micrometer-sized atomically flat terraces (supplementary figure S1) that serve as an ideal surface for molecular adsorption. The samples were then ex-situ transferred to the CA growth chamber for growth of Au//CA films. To obtain a Co surface, 15 nm of Co were thermally evaporated onto these mica/Au substrates at room temperature in a separate UHV chamber. Then the CA thin films were deposited in the same chamber. Polarization studies including ferroelectric domain imaging and local polarization switching were performed ex-situ using a commercial Atomic Force Microscope (Bruker Icon AFM) in PFM mode on freshly prepared samples. The domain imaging was performed by applying an alternating excitation signal of $\pm 4V$ at a frequency of 15 kHz. Local polarization switching experiments were carried out by applying a voltage ramp from -12V to 12V at a speed of 0.2 V/s in addition to the excitation signal.

We first discuss the growth and ferroelectric domain structures of a 50 nm-thick CA film grown on Au. The surface topography and PFM phase signal for out-of-plane/vertical (VPFM) and in-plane/lateral (LPFM) directions of a 2μm×2μm region are shown in figure 1 for two different CA deposition flux rates (0.9 nm/min and 1.6 nm/min). As we see in figure 1, there seems to be a significant effect of the deposition flux rate not only on the surface morphology but also on the ferroelectric domains. By increasing the flux rate from 0.9 nm/min (upper panel) to 1.6 nm/min (lower panel), we observe a reduction in grain size distribution while the RMS roughness decreases from 12.8 nm to 8.7 nm.

The CA deposition flux rate significantly alters the properties of the resulting ferroelectric domains. For a rate of 0.9 nm/min, the similar contrast between VPFM and LPFM maps (figure. 1 b-c) indicates that the polarization is canted with respect to the substrate surface. However, for a rate of 1.6 nm/min, the LPFM phase contrast is easily visible (figure. 1f), while the VPFM phase contrast is clearly minimized (figure 1e). We infer that the polarization is now mainly parallel to the surface. This suggests that the flux rate influences not only the film morphology, but also the structural ordering of these polar molecules.

We present, in Figure 2, a similar study conducted at a fixed 1.6 nm/min flux rate on 50 nm-, 100 nm- and 200 nm-thick CA films. We do not observe a VPFM phase contrast (figure 2b, e and h), which indicates that the ferroelectric polarization remains in-plane in this thickness range for this flux rate. As the thickness increases, the grains become more elongated (figure 2a, d and g). The in-plane ferroelectric domains (figure 2c, f and i) follow the same trend.

To better understand the correlation between the grains and the domains, we extracted the contours of the grains from the topographical images (figure 2a, d and g) and superimposed them on the corresponding phase images (figure 2c, f and i). In this way, the grain boundaries are easily distinguished from domain boundaries. The results are shown in figure 3, where grain boundaries are represented by black lines and the LPFM phase contrast by a white-purple contrast. It can be seen that some domains expand over several grains. More often, the domain boundaries coincide with the grain boundaries and in that case it is difficult to identify the domain walls. But there are domains which are formed within single grains and the boundary of such domains can be identified as domain walls. Such multiple domains with opposite polarization exist within a single grain to lower the electrostatic energy arising due to polarization charges. Some of these multi-domain grains are marked with red dashed circles in figure 3c, most of which possess stripe shaped domain patterns. Such a stripe domain arrangement indicates that a parallel alignment of the polarization with respect to the domain walls is favored.

We then attempted to perform local polarization switching of Au//CA by applying a DC ramp voltage to the PFM tip along with the alternating AC bias. No robust switching of the local polarization was observed on any of the Au//CA samples shown above. We often ended up in distorting the grains and creating morphological voids without any sign of switching either in the LPFM or VPFM phase. An example of such a distortion is shown in figure S4 of the supplementary information.

We now turn to CA films grown on Au//Co. We present, in figure. 4, morphology and PFM phase maps of 50 nm-thick CA films grown at 1.6 nm/min. When deposited on Co, the CA grains adopt a rather elongated shape in contrast to circularly shaped grains on Au for 50 nm-thick films. For example, in figure 4a and 4d, highly directional growth of CA grains is visible, while the film's surface coverage has decreased. The ferroelectric domains follow the morphology and are also likewise elongated. Although the contrast for VPFM is not as sharp as it is for LPFM, the presence of a non-zero VPFM contrast indicates an out-of-plane component of the polarization vector for some grains.

In contrast to the case of CA films grown on Au, local polarization switching of a single grain of a CA film grown on Co was successfully observed. We present, in figure 5a, b and c, hysteresis loops of three different PFM quantities, namely, VPFM, LPFM and relative amplitude of tip deflection. These data show that a polarization switching is possible at a moderate voltage. We observe that the reversal curve/hysteresis of the LPFM phase (figure 5b) is much sharper than that of VPFM phase (figure 5a), as expected since the polarization state is in-plane (figure 4). Nevertheless, tiny switching features can also be noticed in the VPFM phase reversal curve (gray arrows of figure 5a) that appear at exactly the same voltage values as those for the LPFM phase reversal curve. The successful phase reversal of vertical and lateral phases confirms the ferroelectricity of CA on Co.

The lower panels of figure 5 demonstrate a successful domain writing process. Figure 5c and figure 5g show the initial FE state as imaged using VPFM and LPFM, respectively. The phase images after poling the domain with +12 V and -12 V are shown in figure 5e and 5f, respectively, for the VPFM phase and in figure 5h and 5i, respectively, for the LPFM phase. A clear change of phase contrast between the two oppositely polarized states is evident in both VPFM and LPFM phases for the grain of interest (dashed circles).

The polarization reversal observed in figure 5 is happening mainly in the in-plane direction. But the minor switching features in VPFM phase reversal curve and a clear change in VPFM contrast of the domain indicate the presence of a non-zero component of polarization reversal occurring in the out-of-plane direction as well. Furthermore, it can be noted that a field applied perpendicularly to the surface is able to reverse the in-plane polarization which is indicative of coupling between both polarization components. These two facts indicate that the net polarization axis is canted with respect to the film plane at a small angle.

We present, in figure 6, spatially detailed LPFM maps acquired after consecutive polarization reversal experiments on a 50 nm-thick CA film on Co. Repeated poling of the grain (white circle) resulted in the systematic reversal of the polarization state of a part of the grain, with the voltage polarity determining the resulting polarization state.

Density Functional Theory and Molecular Dynamics calculations predict that it should be possible to switch the polarization in a 2D layer of CA on the surfaces of Au and also on more reactive surface of Ag.[27,28,33] However, the experimental confirmation of such switching at a few monolayer level is not easy and has been lacking so far. In 2D layers, the polarization of the CA network is essentially in the plane of the film requiring an in plane electric field for switching, whereas, in a bulk polycrystalline film the polarization axis can be at an angle with respect to the film plane. So even with an out of plane electric field like that emanating from a PFM tip, it is expected that polarization reversal can take place both in and out of the plane directions. Thus it is unclear why on Au surface in bulk films reversible ferroelectric properties of CA are not robust whereas on Co surface they are. In the absence of any structural information of the CA grains on both the surfaces, we attempt to qualitatively explain the anomaly as the following.

Recent STM studies on 2D layer growth of CA on Au and Ag surfaces show that CA molecules form different types of networks depending on the substrate, primarily due to different substrate-molecule interactions.[27,28] Calculations show that the interaction of CA with Au is weaker than that with Ag for Ag is a more reactive substrate. Further, STM studies reveal high surface mobility and low diffusion barrier of CA layers on Au surface. Highly unstable networks of CA for less than 1 monolayer thickness coverage have also been observed on Au surface.[28] Similar studies of CA on Co surface have not been reported so far but thanks to higher surface reactivity /Co-molecule interaction on Co surface, stable networks of CA layers can be expected. This is consistent with our observation of robust properties of CA on Co surface. We believe that the weaker substrate molecule interaction in the initial layers of CA growth on Au surface is at the origin of the observed lack of robustness in the ferroelectric properties on Au surface. This may also account for the observed distortion of the film topography after attempts to switch the polarization in the case of an Au substrate. Thus, we observe that metallic substrates like Co with relatively higher surface reactivity is helping to stabilize the CA ferroelectric properties. This, however, is in contrast to the usually observed scenario where reducing the surface reactivity stabilizes the molecular properties.

On Au surface, the presence of multiple domains in a single grain of CA is a sign of ferroelectric order being present but the lack of robust polarization reversal raises doubts on the ease of switchability of CA grains on Au surface. As the proof of ferroelectricity necessarily requires the observation of polarization reversal, it is difficult to conclude on the ferroelectric response of CA on Au surface, however, the piezoelectric response is evident from the presence of sharp contrast in the vertical and lateral phase images. On Co surface, switching is happening predominantly along the film plane with a small component along the normal to it.

In conclusion, it appears that it is possible to fabricate switchable and robust ferroelectric Croconic Acid thin films on Co, whereas, on less reactive substrates like Au it is not so obvious. Our observations could work as a guide for the development of multifunctional devices based on ferroelectric Croconic Acid composite films. The stability of ferroelectric properties on Co surface combined with the high polarization charge density of CA should encourage fabrication of devices for spintronics applications; for example, organic (ferroelectric)-inorganic (ferromagnetic) hybrid artificial multiferroic devices, organic ferroelectric tunnel junctions (O-FTJ) etc. Moreover, our work also hints at possible rich interface phenomena that may be occurring at a ferromagnetic/organic ferroelectric interface and its role in the stabilization of polarization reversal which is not in the scope of the present work.

Acknowledgements: We are thankful to our funding agency, ANR and DFG for funding the project, ORINSPIN (ANR-16-CE92-005-01). We thank Prof. Dr. W. Wulfhekel of KIT, Germany and Dr. S. Cherifi-Hertel of IPCMS, CNRS-University of Strasbourg, France for their valuable advice and suggestions. We are thankful for the assistance from Mr G. Schmerber and Mr C. Kieber of IPCMS, CNRS in carrying out auxiliary tests and experiments. We acknowledge the assistance from Mr C. Roumegou with some image processing. Lastly but more importantly, we honor the efforts of the late Dr. Eric Beaurepaire of IPCMS, CNRS in designing the project.

# Figures:

Figure 1:

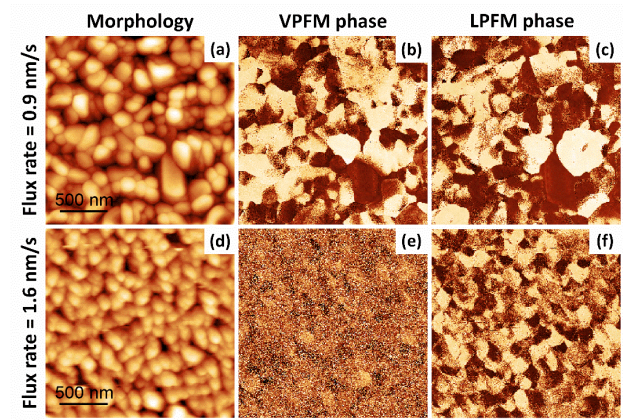

**Figure 1: Flux rate dependence of morphology and ferroelectric domain structure for Au//CA (50nm).** (a, d) Surface morphology, (b, e) out-of-plane PFM phase, and (c, f) in-plane PFM phase of the same 2x2μm² sample region. The flux rate rates were 0.9 nm/min (panels a-b-c) and 1.6 nm/min (panels d-e-f). Scaling is normalized to minimum and maximum values: (a)=(0 nm,98 nm), (b)=(0°,252°), (c)=(0°,326°), (d)=(0 nm, 61 nm), (e)=(0°,241°), (f)=(0°,249°). The PFM amplitude images are shown in supplementary figure S2.

Figure 2:

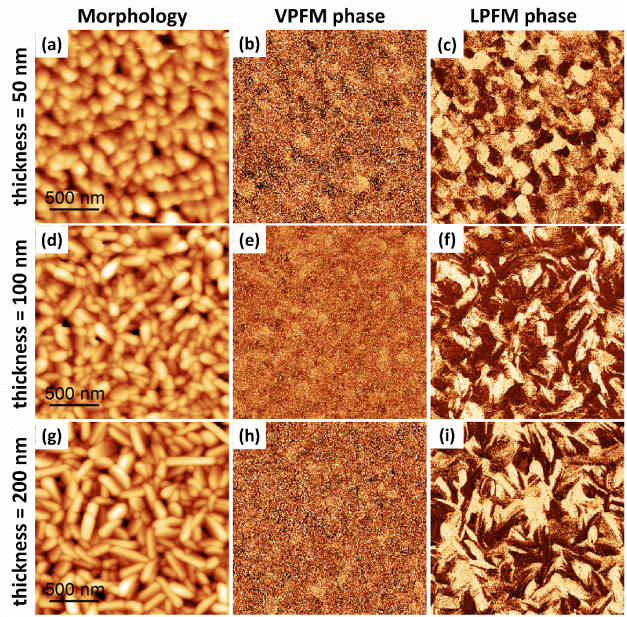

**Figure 2: Thickness dependence of morphology and ferroelectric domain structure of Au//CA grown at a flux rate of 1.6 nm/min.** Upper, middle and lower panels show respectively the topography, out of plane PFM phase and in plane PFM phase for film thickness (t) of 50 nm (upper panels), 100 nm (middle panels), 200 nm (lower panels). For easy comparison, the upper panel is kept the same as the lower panel of figure 1. Scaling is normalized to minimum and maximum values: (a)=(0 nm,61 nm), (b)=(0°,240°), (c)=(0°,324°), (d)=(0 nm, 84 nm), (e)=(0°,140°), (f)=(0°,268°), (g)=(0 nm, 79 nm), (h)=(0°,170°), (i)=(0°,217°). The PFM amplitude images are presented in supplementary figure S3.

Figure 3:

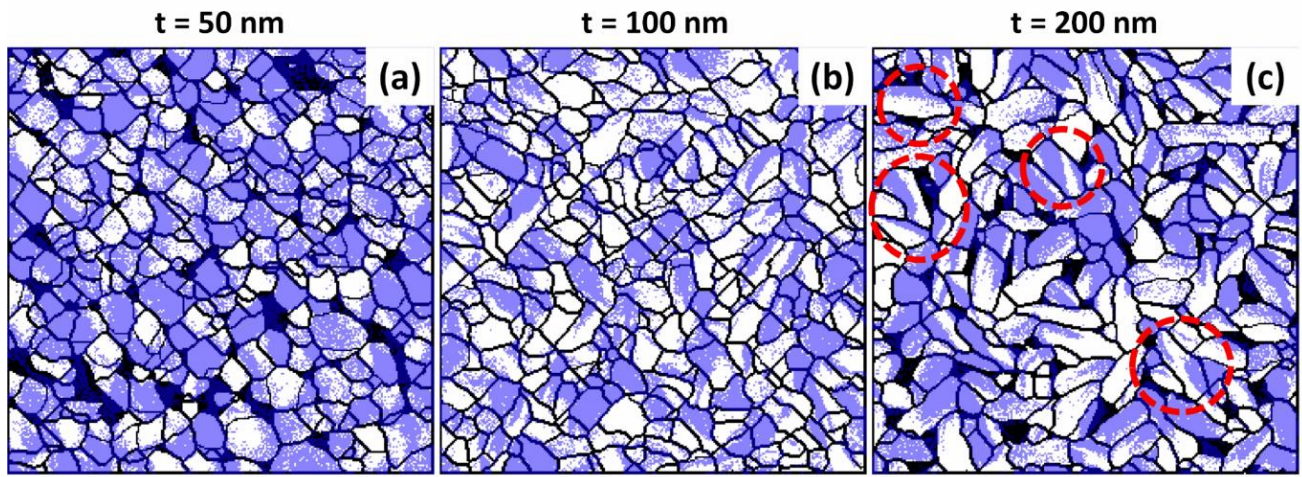

**Figure 3: Topography contours superimposed on LPFM phase maps.** Black lines represent the grain boundaries and the white (minimum)-purple (maximum) shades represent the contrast of LPFM phase response of Au//CA films of thickness, t (a) 50 nm, (b) 100 nm and (c) 200 nm in a region of area 2x2μm$^2$. Red dashed circles in panel (c) show some multi-domain grains. Black regions represent morphological voids.

Figure 4:

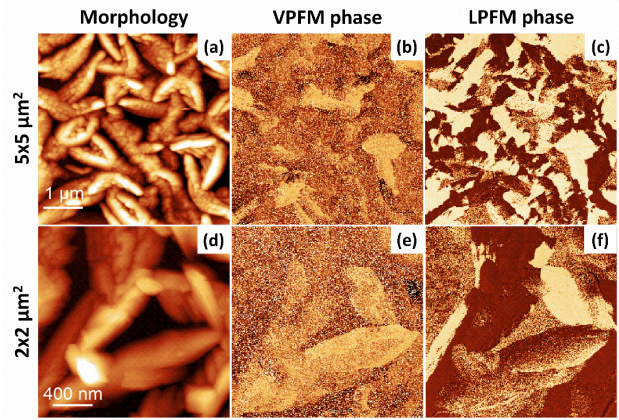

**Figure 4: Surface morphology and ferroelectric domains of 50 nm of Co/CA.** Morphology (a, d), VPFM phase (b, e) and LPFM phase (c, f) are shown with large (upper panel) and small (lower panel) scales. The scales for phase images correspond to the scales shown in the morphology in respective panels. Scaling is normalized to minimum and maximum values: (a)=(0 nm,136 nm), (b)=(0°,226°), (c)=(0°,360°), (d)=(0 nm, 216 nm), (e)=(0°,263°), (f)=(0°,331°). The PFM amplitude images are presented in supplementary figure S5.

Figure 5:

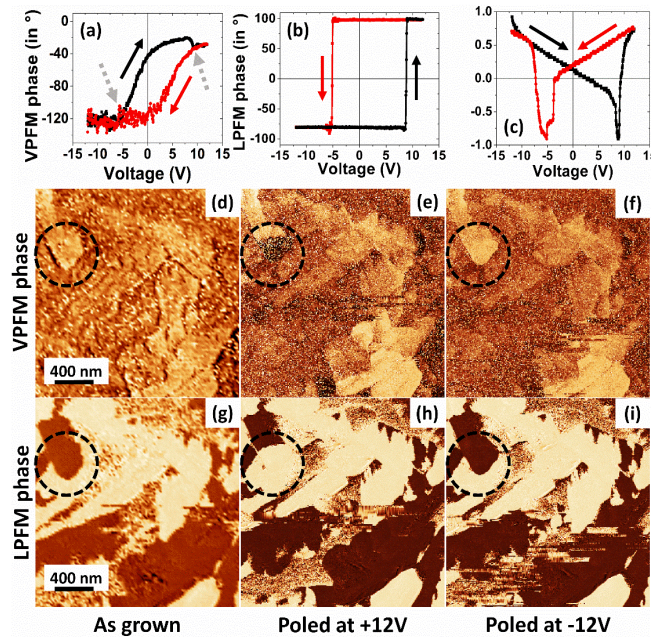

**Figure 5: Polarization reversal hysteresis loops and phase maps of poled states of a domain in Co/CA film.** Upper panel shows polarization reversal curves in terms of VPFM phase (a), LPFM phase (b) and relative grain deformation(c). Dotted arrows in (a) indicate the weakly visible reversal features. Red and black curves represent voltage ramp directions, red for +12V to -12V and black for -12V to +12V. Middle panel shows the VPFM phase maps in the as grown state (d), after poling with +12 V (e) and after poling with -12 V (f). Lower panels show the LPFM phase maps on the same area as (d), (e) and (f). The as grown states of the domain is shown in (d) and (g) where the dashed circles encircle the grain of interest.

Figure 6:

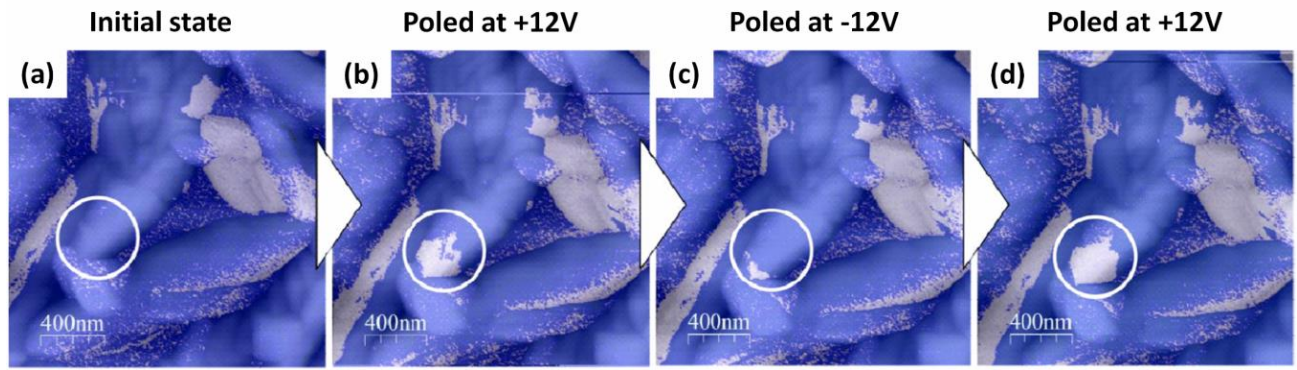

**Figure 6: Robust polarization reversibility of CA domain on Co surface.** Grey-blue contrast shows in-plane domains in the LPFM phase maps of a 2x2μm² region in the initial state (a), after consecutive poling at +12V (b), -12V (c) and at +12V (d).

# Supplementary material for
# Robust ferroelectric properties of organic Croconic Acid films grown on spintronically relevant

Sambit Mohapatra[a*], Victor Da Costa[a], Garen Avedissian[a], Jacek Arabski[a], Wolfgang Weber[a], Martin Bowen[a], Samy Boukari[a]

[a] Université de Strasbourg, CNRS, Institut de Physique et Chimie des Matériaux de Strasbourg, UMR 7504, F-67000 Strasbourg, France

*Correspondence E-mail: sambit.mohapatra@ipcms.unistra.fr

**Supplementary figures:**

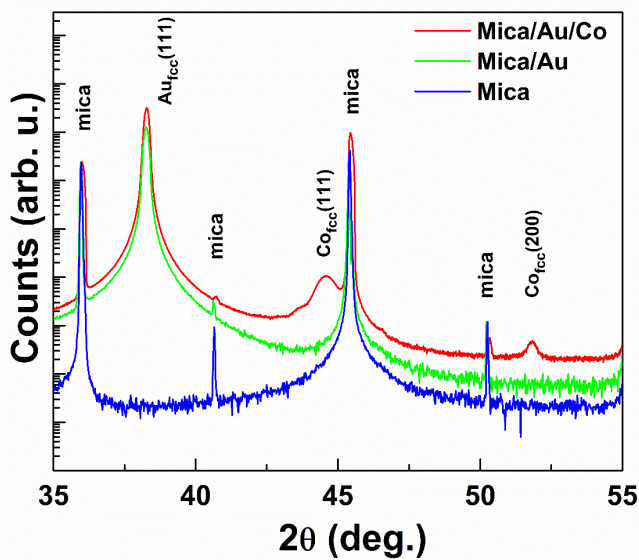

*Figure S1:* XRD of mica (blue), mica//Au (black) and mica//Au/Co (red).

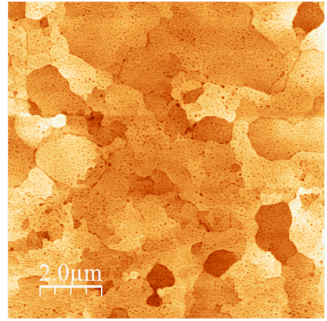

*Figure S2:* Surface morphology of mica/Au substrate showing micrometer sized flat terraces.

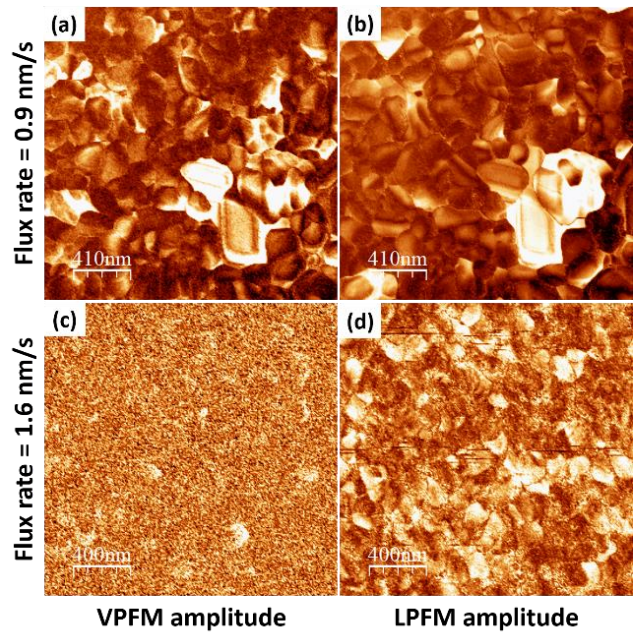

*Figure S3:* Flux rate dependence of PFM amplitude for Au//CA (50nm) film, corresponding to figure1 in the main text.

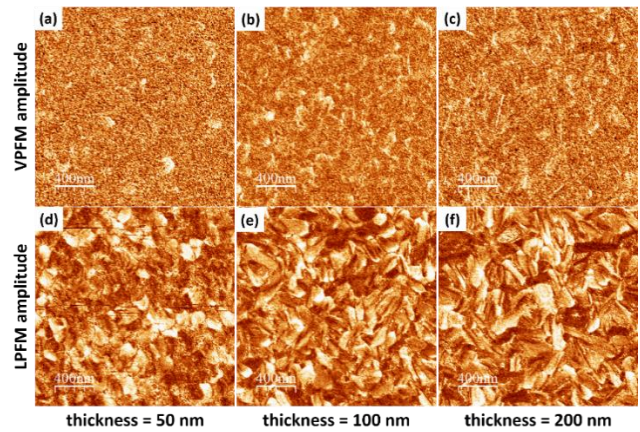

*Figure S4:* Thickness dependence of PFM amplitudes for a growth rate of 1.6nm/min on Au//CA, corresponding to figure2 in the main text.

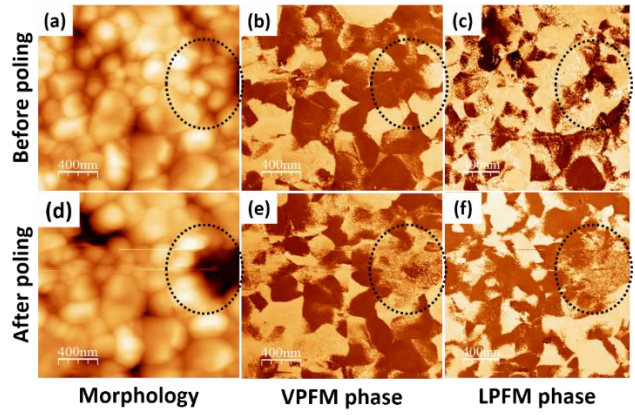

*Figure S5: Impact of poling bias on the morphology of Au//CA (50 nm) sample.*

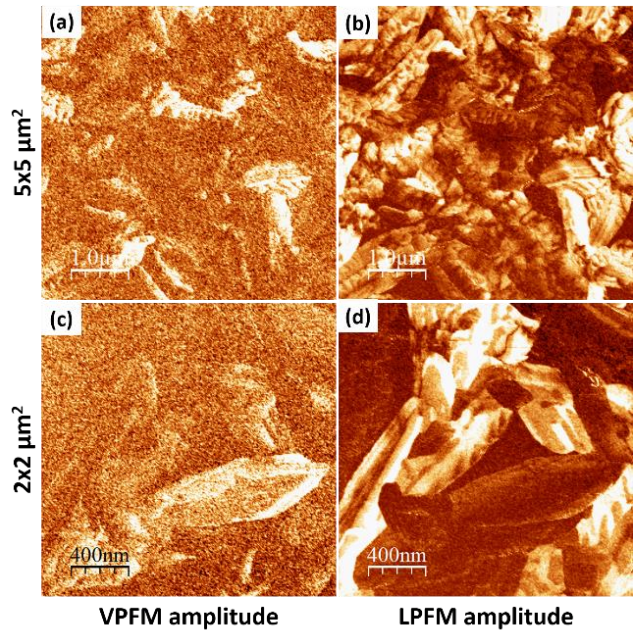

*Figure S6: PFM amplitude corresponding to figure 4 in the main text.*

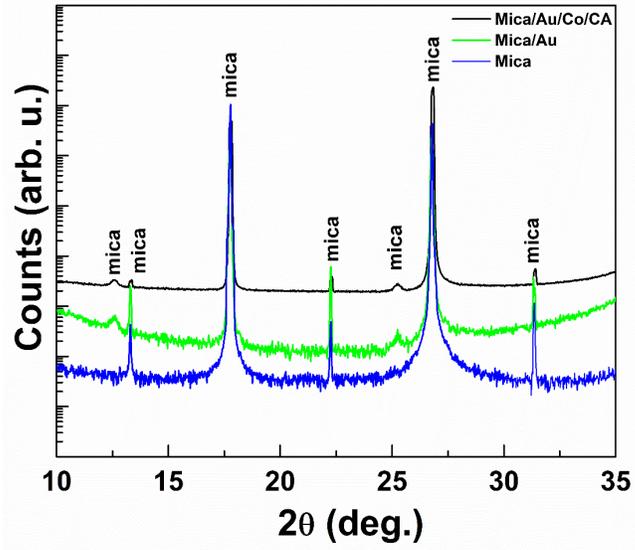

*Figure S7:* XRD of mica (blue), mica//Au (green) and mica//Au/Co/CA (black).

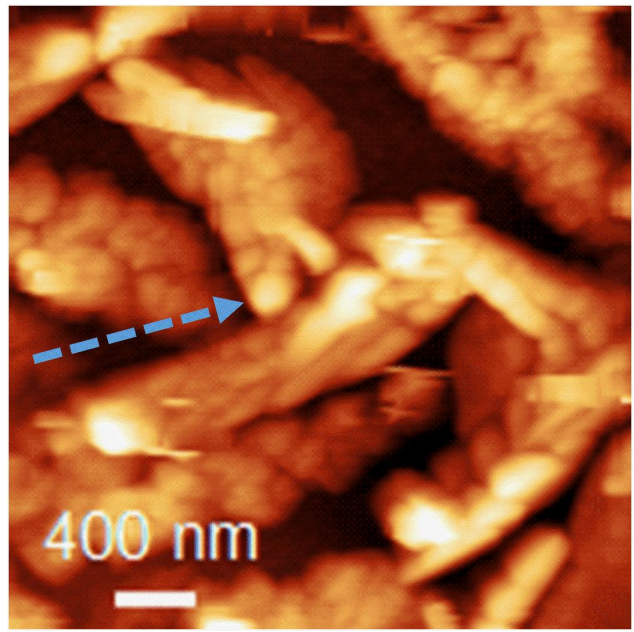

*Figure S8:* Surface morphology of the grain for which switching is shown in figure 5 in the main text.